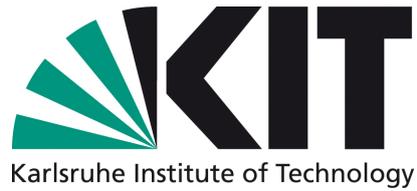

Karlsruhe Institute of Technology

**This is the author's version of a work that is published through the following outlet:**

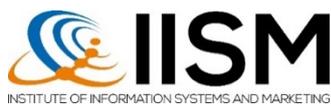

**Institute of Information Systems and Marketing (IISM)**
Kaiserstr. 89
76133 Karlsruhe
Germany

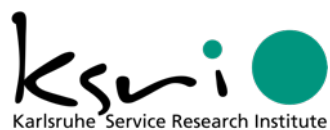

**Karlsruhe Service Research Institute (KSRI)**
Kaiserstr. 89
76133 Karlsruhe
Germany

# UNDERSTANDING DATA UNDERSTANDING: A FRAMEWORK TO NAVIGATE THE INTRICACIES OF DATA ANALYTICS

*Completed Research Paper*

Joshua Holstein, Karlsruhe Institute of Technology, Karlsruhe, Germany, joshua.holstein@kit.edu

Philipp Spitzer, Karlsruhe Institute of Technology, Karlsruhe, Germany, philipp.spitzer@kit.edu

Marieke Hoell, Karlsruhe Institute of Technology, Karlsruhe, Germany, marieke.hoell@kit.edu

Michael Vössing, Karlsruhe Institute of Technology, Karlsruhe, Germany, michael.voessing@kit.edu

Niklas Kühl, University of Bayreuth, Bayreuth, Germany, kuehl@uni-bayreuth.de

## Abstract

*As organizations face the challenges of processing exponentially growing data volumes, their reliance on analytics to unlock value from this data has intensified. However, the intricacies of big data, such as its extensive feature sets, pose significant challenges. A crucial step in leveraging this data for insightful analysis is an in-depth understanding of both the data and its domain. Yet, existing literature presents a fragmented picture of what comprises an effective understanding of data and domain, varying significantly in depth and focus. To address this research gap, we conduct a systematic literature review, aiming to delineate the dimensions of data understanding. We identify five dimensions: Foundations, Collection & Selection, Contextualization & Integration, Exploration & Discovery, and Insights. These dimensions collectively form a comprehensive framework for data understanding, providing guidance for organizations seeking meaningful insights from complex datasets. This study synthesizes the current state of knowledge and lays the groundwork for further exploration.*

## 1    Introduction

In today's digital age, organizations continuously create and store large amounts of data for analysis (Fassnacht et al., 2023). To extract value from this accumulated data, they increasingly rely on analytics approaches ranging from business intelligence through data analytics to artificial intelligence (AI) (Mikalef et al., 2018; Savarimuthu et al., 2023). However, analyzing real-world data to uncover novel insights can be complex and challenging—it can contain intricacies, biases, and nuances that may be easily overlooked (Holstein et al., 2023). As big data continues to increase in size and diversity, these challenges even amplify (Savarimuthu et al., 2023). It is no longer just a matter of acquiring large amounts of data but of identifying whether the data accurately represents the real-world (Aaltonen et al., 2023; Lebovitz, Levina, and Lifshitz-Assaf, 2021) and which parts of that data are most relevant to specific problems (Holstein et al., 2023). A critical task, therefore, lies in identifying the most relevant subsets or variables within the data, which necessitates an in-depth understanding of its context, quality, and relevance (Gerhart, Torres, and Giddens, 2023). Such an understanding ensures that the correct data is leveraged, enabling organizations to derive accurate and contextually relevant insights from the vast complexity of big data (Abbasi, Sarker, and Chiang, 2016). This understanding increases the efficiency of subsequent data analysis and aligns it with business goals, ensuring actionable insights aligned with organizational objectives.

Despite the established importance of data understanding, many existing analytics frameworks, including the widely adopted Cross-Industry Standard Process for Data Mining (CRISP-DM) (Wirth and Hipp, 2000) or the Knowledge Discovery in Databases (KDD) process (Fayyad, Piatetsky-Shapiro, and Smyth, 1996), cover the data understanding phase only superficially. For instance, CRISP-DM defines data understanding as an early phase of a project, where the goal is to gain a broad understanding of the data's structure, quality, and content (Wirth and Hipp, 2000). However, it does not guide how to achieve this understanding or how it should inform subsequent process phases. Furthermore, it does not explicitly address the complexities of understanding large and diverse datasets, which are common in today's data-driven environments. Simultaneously, the rise of data-centric AI (DCAI) emphasizes that "the systematic design and engineering of data are essential for building effective and efficient AI-based systems" (Jakubik et al., 2024, p. 2). This shift places less emphasis on developing increasingly complex algorithms and more on understanding and improving the data that fuels these algorithms (Whang et al., 2023). A critical aspect of DCAI is the augmentation of data, which involves selecting, modifying, or adding the data to improve its quality for a particular task. This process requires domain knowledge and a deep understanding of the underlying data, as it informs the strategies used to augment the data (Jakubik et al., 2024). Without a thorough understanding of the data, organizations may struggle to effectively augment their data, potentially undermining the performance of their AI systems. This shift towards a data-centric perspective in AI reflects the changing demands in information systems (IS), where data has become increasingly viewed as a critical asset that requires in-depth understanding (Aaltonen et al., 2023; Fassnacht et al., 2023). The lack of guidance on data understanding in existing frameworks, despite it's increasing importance in emerging fields like DCAI, underscores a critical research gap. Therefore, we formulate the following research question:

**RQ**: What are the core dimensions of data understanding that facilitate the extraction of insights from data through analytics?

To address this research question, we explore the role of data understanding as guided by established analytics frameworks. Accordingly, we conduct a systematic literature review aligned with the methodology of Webster and Watson (2002) to perform a qualitative content analysis according to Gioia, Corley, and Hamilton (2013) or Wolfswinkel, Furtmueller, and Wilderom (2013), focusing on high-impact sources within the IS and adjacent computer science disciplines. Our analysis identifies five core dimensions of data understanding: Foundations, Collection & Selection, Exploration & Discovery, Contextualization & Integration, and finally, Insights. These dimensions collectively form a framework, shedding light on the complex process of transforming raw data into sufficiently understood data, making it actionable for analytics like business intelligence or DCAI. Thus, our study contributes to a deeper understanding of both the concept and the processes of data understanding in the IS field by complementing existing analytics frameworks with a holistic view of data understanding. Further, we contextualize our findings on data understanding within the emerging DCAI paradigm and underline its contribution.

The paper is structured as follows: Chapter 2 reviews the literature on data understanding and DCAI development. Chapter 3 outlines our systematic literature review method. Chapter 4 details identified core dimensions and Chapter 5 introduces a framework based upon them Chapter 6 discusses the study's implications and future research directions.

## 2      Background

Due to the rapidly increasing volume and velocity of data, organizations face the challenge of effectively using large and complex data sets to address real-world problems (Savarimuthu et al., 2023). In response to this challenge, various data analytics frameworks have been developed, offering structured approaches to data analysis (Haertel et al., 2022). These frameworks aim to guide users through phases from understanding a business problem to deploying data-driven solutions. However, the treatment of each stage, particularly the initial understanding and integration of data into the analytical process, varies significantly across different frameworks (Haertel et al., 2022). For instance, the KDD process (Fayyad, Piatetsky-Shapiro, and Smyth, 1996) does not have a specific phase for data understanding. Instead, it integrates this activity into several other phases, such as "Creating a target dataset", "Data cleaning and preprocessing", and "Data reduction and projection", emphasizing the preprocessing of the data rather than the underlying understanding. Conversely, while the CRISP-DM model (Wirth and Hipp, 2000) includes a phase for understanding data, its guidance remains limited. Rather than promoting a holistic understanding of data, it merely mentions collecting and describing data before exploring it and verifying its quality for reporting. However, this approach overlooks the crucial role of domain knowledge in comprehensively contextualizing and understanding data (Gerhart, Torres, and Giddens, 2023). Other frameworks follow a similar trend. Many studies emphasize preprocessing and modeling over deep data understanding, for example, Cao et al. (2010), K. Chen and Liu (2006), Dag et al. (2016), and Fayyad, Piatetsky-Shapiro, and Smyth (1996). In addition, existing comparative studies, such as Fatima et al. (2020), Haertel et al. (2022), and Mariscal, Marbán, and Fernández (2010), contrast entire frameworks rather than specific phases, such as data understanding. This leaves a gap in our knowledge of what constitutes data understanding in the context of analytics projects. Yet, previous research has highlighted the need for a deeper understanding of data and how it represents the real world (Aaltonen et al., 2023) and the integration of domain expertise (Gerhart, Torres, and Giddens, 2023). This recognized need, combined with the current coverage of data understanding in existing frameworks, underscores the shortcomings of existing frameworks.

In this landscape, the growth of DCAI marks a paradigm shift, emphasizing the importance of systematic design and engineering of data as essential elements for building effective and efficient AI-based systems. Unlike model-centric AI, DCAI focuses on improving the quality and quantity of data given a fixed AI model rather than tuning the model itself (Jarrahi, Memariani, and Guha, 2023). This approach emphasizes the importance of domain-specific data augmentation, complemented by the development of methods and semi-automated tools, to accelerate the development of successful AI-based systems (Jakubik et al., 2024). DCAI thus offers a novel perspective on the role of data quality. It argues that more appropriate data can drive performance improvements and that changes in AI model performance metrics can indicate the effectiveness of adjustments in the data. This perspective emphasizes the need for deep data understanding and the importance of maintaining up-to-date data for training effective models, as emphasized by Zha et al. (2023b). Prominent examples such as Jakubik et al. (2024), Jarrahi, Memariani, and Guha (2023), and Zha et al. (2023a) further underscore the importance of domain knowledge in DCAI, emphasizing the need for in-depth analysis when dealing with large, high-dimensional datasets. On the other hand, Patel et al. (2023) illustrates the value of exploratory data analysis in improving data understanding through novel sampling techniques.

This shift towards DCAI stresses the need for guidance on understanding data effectively. Therefore, our research aims to delineate the core dimensions that constitute a thorough understanding of data within analytics frameworks, considering both the business challenges and the insights offered to DCAI.

## 3 Research Methodology

To explore the mechanisms by which analytics frameworks facilitate data understanding, we employ a systematic literature review methodology, adhering to established IS methods (Webster and Watson, 2002), and adopt an inductive, grounded theory-inspired approach (Wolfswinkel, Furtmueller, and Wilderom, 2013) to understand how data understanding is established across different frameworks. Although synthesizing elements across frameworks may shift focus from their contextual specifics, our goal is to identify common elements that define a holistic data understanding. We follow the recommendations of Gioia, Corley, and Hamilton (2013) to articulate the results of our inductive data analysis.

Before describing the specifics of our systematic review, we clarify its scope using Cooper (1988) taxonomy of literature reviews. The primary aim of our review is to systematize research theories and methodologies. Our objective is to synthesize the existing body of research to identify and elucidate key themes and dimensions. We aim for a neutral perspective and seek to provide a literature review that is representative of the broad connections to other research fields inherent in our topic. Employing grounded theory-inspired methods, we aim for a conceptual synthesis of studies, addressing a general scholarly audience. As suggested by Wolfswinkel, Furtmueller, and Wilderom (2013), we follow the four phases for systematic literature reviews: *define*, *search*, *select*, *analyze*, and *present*.

**Define.** Similar to Raftopoulos and Hamari (2023), we review an initial set of frameworks to familiarize ourselves with the concept of data understanding and its role in data analytics projects. This allowed us to develop a shared understanding of the underlying phenomena, thus defining the inclusion and exclusion criteria and developing our search terms. Following our research question, we define criteria to include articles that present a framework for analytics, i.e., conceptual papers or official documentation introducing specific frameworks. We select official documentation based on forward and backward search. Further, discuss how they build an initial data understanding of the underlying data or articles that offer an interpretation, discussion, expansion or comparison of frameworks or associated challenges. The following questions guided our search:

- How do analytics frameworks define the stage of data understanding?

- How do these frameworks address the complexity and variety of data to generate insights?

- What is the outcome of data understanding?

We excluded articles that have their primary contribution in (automated) technical methods, for example, in augmenting or cleaning datasets, rather than methodological guidance on obtaining data understanding in analytics projects. Additionally, we exclude all articles that do not mention concepts and activities related to data understanding or are in a language other than English. Furthermore, we identified several relevant synonyms for data analytics, which we included in our search term in an iterative search and refined the process, ultimately resulting in *("data scien\*" OR "data mining" OR "data analytics" OR "big data" OR "knowledge discovery" OR "data analysis") AND ("process model" OR "framework" OR "methodology") OR ("data understanding")*. As data analytics projects are not only relevant to IS researchers, we also include outlets in the adjacent field of computer science. Our research focuses on sources with the highest impact, specifically selecting outlets from the senior scholars' list of premier journals in the IS discipline and journals ranked as A\* or A according to the CORE ranking. This selection is tailored to include publications related to IS or data mining, i.e., IEEE Transactions on Knowledge and Data Engineering, Data Mining and Knowledge Discovery, IS, ACM Transactions on IS, and ACM Transactions on Database Systems, as we consider these outlets to be a representative sample for high-quality research in the discipline of data analytics frameworks in the fields of IS and computer science. We conduct our search in the interdisciplinary databases Web of Science, Scopus, and the AIS library to collect potentially relevant articles. As data analytics is a research field with a long history and many standards established early, we include all articles published after January 1, 1995 until September 1, 2023.

**Search.** After several rounds of refining the search terms, we conduct a final literature search in the databases mentioned, which results in a literature sample of 1340 articles.

**Select.** Based on our initial search, we first remove duplicates, narrowing our sample down to 808 articles. Afterward, we perform an initial screening by reading the title and abstract (Snyder, 2019) using the inclusion and exclusion criteria. After the screening, we include 40 articles for our full-text screening, thereby narrowing our dataset down to 21 relevant papers. Finally, we perform a forward and backward search (Webster and Watson, 2002) on this sample, thus identifying 17 additional articles. In this final stage, our literature sample includes 38 articles that introduce frameworks for data analytics-related projects mentioning data understanding.

**Analyze.** During the analysis of the selected paper, we follow the methodological guidelines outlined by Wolfswinkel, Furtmueller, and Wilderom (2013), using an iterative coding scheme consisting of three parts: Open, axial, and selective coding. As per the suggestions of Gioia, Corley, and Hamilton (2013) for presenting coding results, we utilize a hierarchical order to describe emerging categories, employing the terms "concept," "theme," and "dimension". Initially, we carefully study the articles and record essential bibliographical details, including the year of publication and the publication outlet. Simultaneously, we extract excerpts that address our research question for annotation of the open coding. Three researchers independently conduct this annotation on a representative subset of the selected paper. After completing the open coding process, we conduct a workshop among the authors to establish a common understanding of the concepts strengthen the inter-coder reliability for subsequent coding activities. Similar to Hund et al. (2021) and Spitzer et al. (2023), one author refines the codes based on the established understanding. In a second workshop, we discuss the refined codes and finalized them. Afterward, we used axial coding techniques to map the interrelationships between first-order concepts and second-order themes. Finally, we used selective coding to build a framework encompassing the identified main categories.

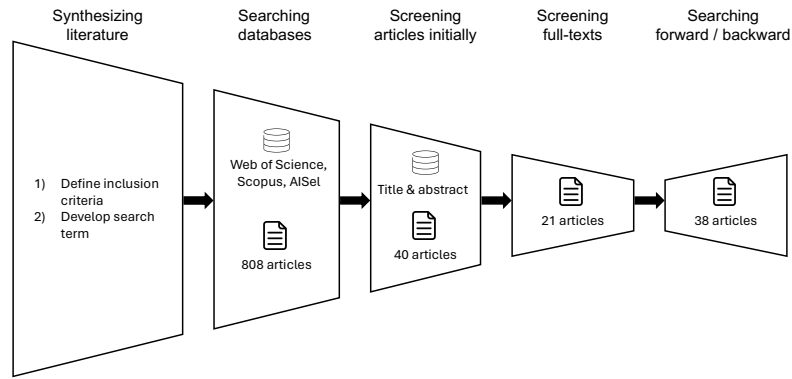

*Figure 1.*          *Steps of the conducted literature search*

# 4      Results

Next, we present dimensions derived from our systematic literature review on data understanding within analytics frameworks. Following open coding of the collected literature to extract first-order concepts, we proceed to axial coding, where these concepts are synthesized into more abstract, second-order themes. Finally, these are aggregated into dimensions that offer a synthesized view, capturing the complex relationships and patterns in data understanding across various frameworks.

## 4.1      Foundations

As the first dimension, we present data foundations as essential to understanding data. This dimension is defined by characterizing the data, understanding transformations that have been applied before storing it in data warehouses, and, finally, the extraction from these warehouses (see Figure 2).

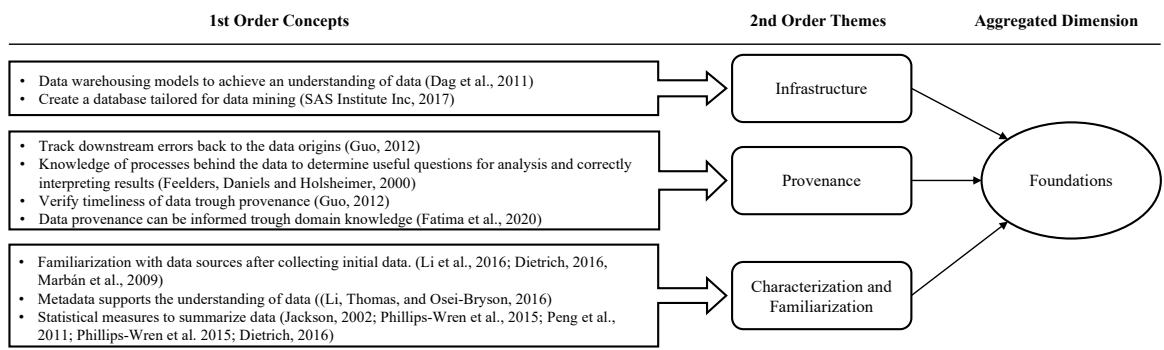

*Figure 2.*          *Data structure of foundations*

**Infrastructure** emphasizes the importance of data warehousing models and tailored databases for data mining. Data warehousing models provide a structured environment for data storage, facilitating efficient data analysis (Dag et al., 2016). Similarly, databases designed for data mining, equipped with summary statistics, prepare data for in-depth analysis (SAS Institute Inc, 2017). These infrastructure elements are crucial for organizing and setting a solid foundation for data exploration and understanding.

**Provenance** represents comprehending the data's origin, the applied transformations, and its timeliness as part of data foundations. The ability to trace data provenance and document the transformation of data is critical to accurately interpreting results and preventing misinterpretations (Feelders, Daniels, and Holsheimer, 2000). Tracing data provenance enables verification that the data is still up-to-date and relevant to the business problem at hand (Guo, 2012). Additionally, it assures a transparent path for the data. It preserves its integrity, enabling the tracing of potential downstream errors back to their origins (Guo, 2012) to adjust and mitigate the root causes of these errors. Understanding the data collection processes and their respective transformations allows for formulating hypotheses that aid in later analyses, identifying relevant data to answer those hypotheses, and potentially creating new features (Feelders, Daniels, and Holsheimer, 2000). The analysis of the data provenance can be informed by the domain knowledge of various experts (Fatima et al., 2020).

**Characterization and Familiarization**, particularly through metadata, is underscored across studies. Metadata serves a critical role by providing detailed descriptions of the data and its linkage to underlying business processes, thereby facilitating a deeper understanding of the data's utility (Li, Thomas, and Osei-Bryson, 2016). This metadata is instrumental in bridging the gap between raw data and its practical application within business environments. To complement available metadata, statistical measures or tools can effectively summarize the data, offering insights into its distribution patterns and underlying structures (Jackson, 2002; Phillips-Wren et al., 2015). Further, examining data features and characteristics can facilitate a detailed understanding of data segments and how they contribute to the broader dataset (Peng et al., 2011). Further, it is important for analysts to familiarize themselves with different data sources, particularly after the initial collection (Li, Thomas, and Osei-Bryson, 2016; Marbán et al., 2009). This process involves thoroughly assessing the data's granularity, aggregation levels, and value range of each data source (Dietrich, 2016), ensuring that the data landscape is comprehensively understood. Such familiarization is a critical data understanding task, pivotal in determining if all key data elements are present or if additional data needs to be sourced (Dietrich, 2016). This comprehensive approach to familiarization aids in making informed decisions about the suitability and completeness of the data for specific analytical tasks, avoiding pitfalls such as over-generalization or inadequate detail in analysis.

## 4.2 Collection and selection

To understand the data, it is essential to collect the right data, i.e., to be able to boil down the potentially large data set to the relevant data and eventually to supplement existing data with new data.

**Data Collection** encompasses the acquisition of data as well as decisions regarding the selection and evaluation of this data for further use. As part of data understanding, the initial data collection sets the foundation for subsequent stages of analysis (Cios and Kurgan, 2005; Haertel et al., 2022; Marbán et al., 2009; Rollins, 2015). A profound knowledge of the data available both within and outside the organization is emphasized as crucial for effective data selection (Feelders, Daniels, and Holsheimer, 2000). This knowledge aids in identifying gaps in the current data landscape and in making informed decisions about which additional data sources might be beneficial for enriching the analysis. One needs to determine whether the data size is appropriate to achieve the underlying goals (Yu, Wang, and Lai, 2006).

**Selecting Relevant Data** might be necessary if the dataset is large and includes instances or variables not relevant to the business use case. Analysts are often presented with extensive datasets, but the real-world value lies in particular subsets. Discriminating between different subsets and identifying those that are worthy of more in-depth analysis is a fundamental aspect of data understanding (Brachman and

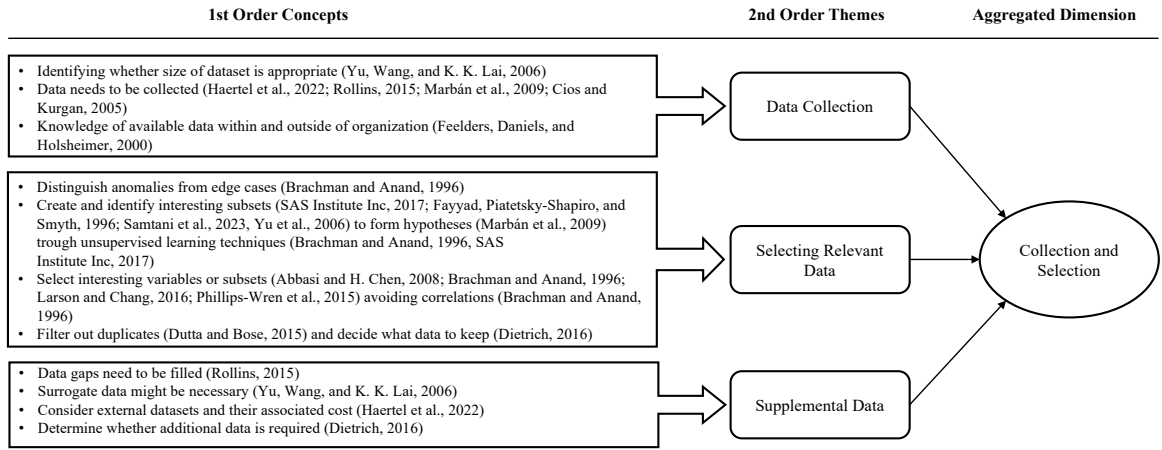

1st Order Concepts — 2nd Order Themes — Aggregated Dimension

*Figure 3.        Data structure of collection and selection*

Anand, 1996; Dutta and Bose, 2015; Fayyad, Piatetsky-Shapiro, and Smyth, 1996; Samtani et al., 2023). This focused analysis enables the formulation of hypotheses based on the insights obtained from these critical subsets, effectively allocating resources and attention to the most informative subsets of the data (Marbán et al., 2009). Different methods can be applied to identify these potentially interesting subsets of the data, including unsupervised learning techniques (Brachman and Anand, 1996; SAS Institute Inc, 2017). Complementing this, data mining methods are employed to discern relevant variables, effectively differentiating between critical data for analysis and extraneous information that can be excluded (Peng et al., 2011). Concurrently, feature selection techniques, such as ranking and projection methods (e.g., correlation, information gain, or principal component analysis), can be applied to refine the dataset further. These techniques support isolating the most promising features, considering correlations and other statistical measures to ensure a focused and efficient analysis (Abbasi and H. Chen, 2008; Brachman and Anand, 1996; Larson and Chang, 2016; Phillips-Wren et al., 2015). This process of selective parameter identification is important as it recognizes that not all variables contribute equally, enabling a more targeted and meaningful analysis based on an understanding of data correlations and redundancies. A part of selecting relevant data involves identifying problems with specific instances. Here, distinguishing outliers from edge cases is an important consideration (Brachman and Anand, 1996). Outliers are data points that significantly differ from other observations and may represent errors in the data. In contrast, edge cases represent instances that, while unusual, are still valid and relevant to the analysis. This distinction helps to ensure that the data used for analysis accurately represents the underlying trends and patterns. Finally, after identifying the relevancy of different features and instances, one must decide what data to keep or discard (Dietrich, 2016), for example, duplicates (Dutta and Bose, 2015).

**Supplemental Data** might be necessary to be collected if not all required data is available. Often, the initial phase of data understanding reveals gaps or areas where additional information is needed to align with the objectives of the underlying project. In such cases, the collection of supplementary data becomes essential. This might involve acquiring data that was not initially considered or delving deeper into specific areas to gain an in-depth understanding (Dietrich, 2016; Rollins, 2015). Further, situations may emerge where surrogate data is necessary (Yu, Wang, and Lai, 2006). Surrogate data refers to alternative or proxy data used when primary data is unavailable or insufficient. This type of data can provide valuable insights and support conclusions, especially in scenarios where obtaining the original data is challenging or impossible (Yu, Wang, and Lai, 2006). Another approach to filling existing data gaps is considering external data sources and their associated costs. In today's data-rich environment, understanding what external data can be leveraged and at what expense is pivotal for enriching the internal datasets and providing a more comprehensive view (Haertel et al., 2022).

## 4.3    Contextualization and integration

The identified articles in our review present multiple methods to contextualize available data to facilitate an in-depth understanding (see Figure 4).

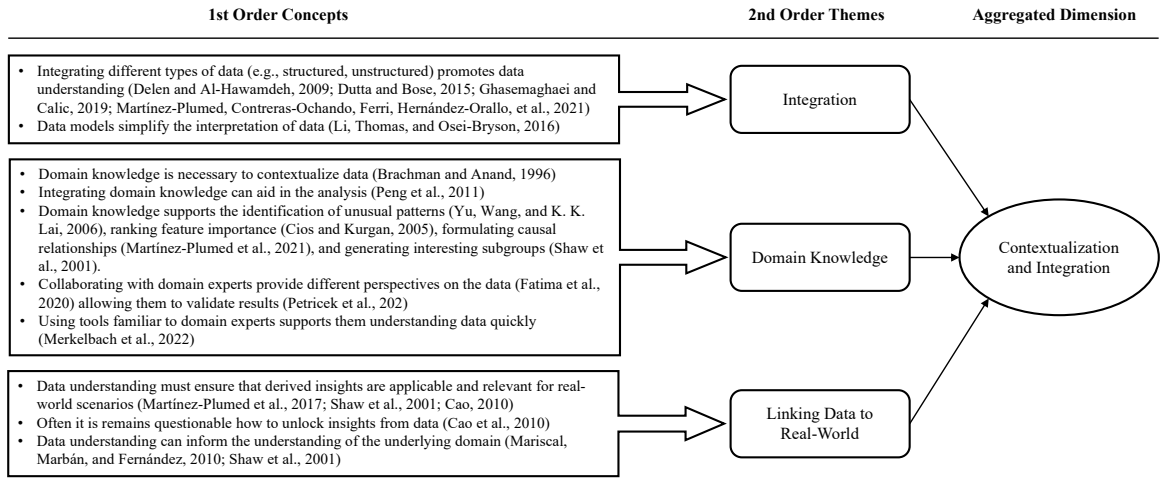

| 1st Order Concepts | 2nd Order Themes | Aggregated Dimension |
|---|---|---|

- Integrating different types of data (e.g., structured, unstructured) promotes data understanding (Delen and Al-Hawamdeh, 2009; Dutta and Bose, 2015; Ghasemaghaei and Calic, 2019; Martínez-Plumed, Contreras-Ochando, Ferri, Hernández-Orallo, et al., 2021)
- Data models simplify the interpretation of data (Li, Thomas, and Osei-Bryson, 2016)

→ **Integration**

- Domain knowledge is necessary to contextualize data (Brachman and Anand, 1996)
- Integrating domain knowledge can aid in the analysis (Peng et al., 2011)
- Domain knowledge supports the identification of unusual patterns (Yu, Wang, and K. K. Lai, 2006), ranking feature importance (Cios and Kurgan, 2005), formulating causal relationships (Martínez-Plumed et al., 2021), and generating interesting subgroups (Shaw et al., 2001).
- Collaborating with domain experts provide different perspectives on the data (Fatima et al., 2020) allowing them to validate results (Petricek et al., 202)
- Using tools familiar to domain experts supports them understanding data quickly (Merkelbach et al., 2022)

→ **Domain Knowledge** → **Contextualization and Integration**

- Data understanding must ensure that derived insights are applicable and relevant for real-world scenarios (Martínez-Plumed et al., 2017; Shaw et al., 2001; Cao, 2010)
- Often it is remains questionable how to unlock insights from data (Cao et al., 2010)
- Data understanding can inform the understanding of the underlying domain (Mariscal, Marbán, and Fernández, 2010; Shaw et al., 2001)

→ **Linking Data to Real-World**

*Figure 4.        Data structure of contextualization and integration*

**Integration** of data sources allows to investigate the interplay of various data sources and types to provide deepening insights. Incorporating different data sources, whether structured or unstructured, can provide a more comprehensive context for analysis. This process facilitates a holistic view that captures the multifaceted nature of data, leading to more informed and accurate insights (Delen and Al-Hawamdeh, 2009; Dutta and Bose, 2015; Ghasemaghaei and Calic, 2019; Martínez-Plumed et al., 2021). The utilization of preexisting data models enhances the understanding of data. Data models function as frameworks that classify and interpret diverse data types, simplifying intricate information architectures (Li, Thomas, and Osei-Bryson, 2016).

**Domain Knowledge** is critical to contextualize and, ultimately, make sense of data. Acquiring domain knowledge is crucial in contextualizing and comprehensively understanding data (Brachman and Anand, 1996). Incorporating acquired domain knowledge can aid in the analysis of data (Peng et al., 2011) by identifying uncommon patterns (Yu, Wang, and Lai, 2006), ranking feature importance (Cios and Kurgan, 2005), formulating causal relationships (Martínez-Plumed et al., 2021), and generating data subgroups (Shaw et al., 2001). Collaborative efforts in data analysis are essential beyond the individual acquisition of domain knowledge. The participation of domain experts provides different perspectives and expertise, leading to a more holistic understanding of the data (Fatima et al., 2020) and allowing them to validate the results of automated analysis (Petricek et al., 2022). Employing tools familiar to domain experts facilitates a quick understanding of the data for domain experts (Merkelbach et al., 2022).

**Linking Data to Real-World** involves interpreting data as well as applying gained knowledge to address and learn about real-world complexities and challenges. Central to this theme is the recognition that real-world data's complexity necessitates thorough analysis to ensure it's applicability and practical relevance (Cao, 2010). In any analytics project, data is seen as a key driver, providing the basis for problem-solving and decision-making. However, it is often not obvious how to unlock its knowledge to provide real-world value (Cao et al., 2010). Through data understanding, one gains insights into the domain, which in turn may inform and refine the approach to data analysis (Cao, 2010; Mariscal, Marbán, and Fernández, 2010). Anticipating the context in which data will be used is also highlighted as critical. It requires activities dedicated to envisaging how data insights will apply in real-world scenarios, ensuring that the analysis remains relevant and grounded in practicality (Martínez-Plumed et al., 2017).

## 4.4 Exploration and discovery

The next dimension encapsulates the critical stage of delving into data to uncover hidden patterns and relationships, for example, through exploration and visualizations (see Figure 5).

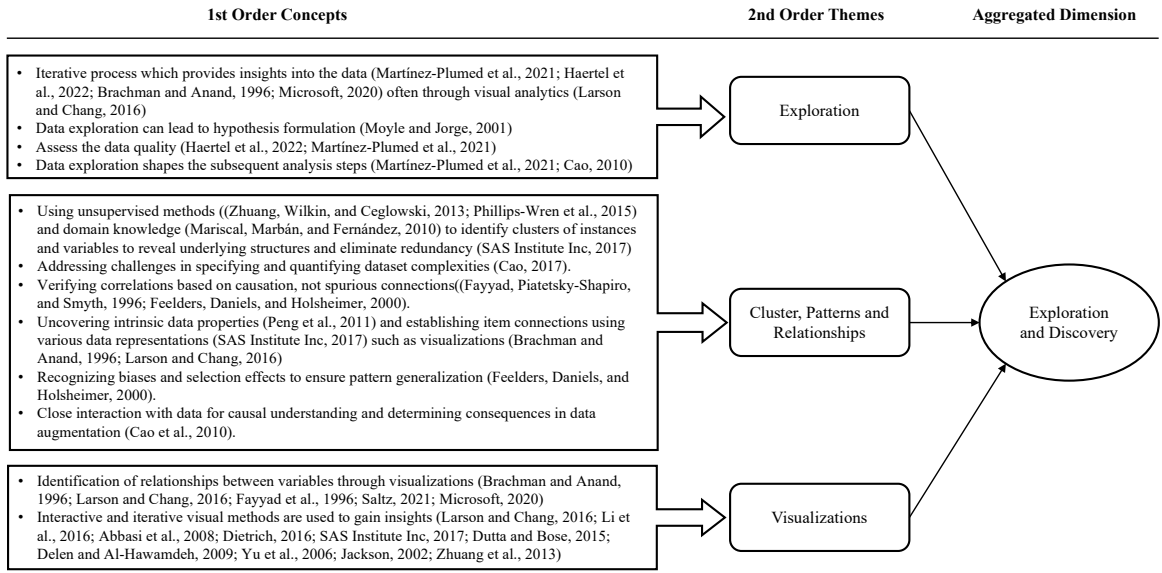

*Figure 5.       Data structure of exploration and discovery*

**Exploration** of data helps to understand and interpret it. It involves iterative engagement with the data, employing various techniques to uncover patterns, relationships, and insights that inform the overall analysis (Brachman and Anand, 1996; Larson and Chang, 2016). Analysts delve into data to observe trends and features, often suggesting new hypotheses about the underlying relationships and phenomena within the data (Moyle and Jorge, 2001). Data exploration also contributes to the development of data descriptions, quality reports, and the understanding of how the data represents its context (Haertel et al., 2022; Martínez-Plumed et al., 2017; Microsoft, 2020). Further, it involves exploring its structure and how it is encoded to inform how the knowledge can be extracted from it (Cao, 2010).

**Clusters, Patterns, and Relationships of Variables** include reducing data complexity by identifying clusters of instances and variables (SAS Institute Inc, 2017), which helps reveal underlying structures and eliminate redundancy as a result of correlated features. Analysts interact closely with the data, requiring a causal understanding to determine consequences in subsequent augmentation of the data (Martínez-Plumed et al., 2021), and verifying identified correlations to ensure they are based on causation rather than spurious connections (Fayyad, Piatetsky-Shapiro, and Smyth, 1996; Feelders, Daniels, and Holsheimer, 2000). Techniques like self-organizing maps (Zhuang, Wilkin, and Ceglowski, 2013) and the integration of domain knowledge (Mariscal, Marbán, and Fernández, 2010) are used to refine pattern discovery, reducing the volume of identified patterns and focusing on relevant insights. The theme also addresses challenges in specifying and quantifying complexities represented in datasets (Cao, 2017) and underscores the importance of visual analytics (Brachman and Anand, 1996; Larson and Chang, 2016) in uncovering variable relationships and presenting interaction information (Abbasi and H. Chen, 2008). Methods such as clustering are instrumental in exploring data and revealing underlying affinity groups (Phillips-Wren et al., 2015). Additionally, recognizing biases and selection effects is crucial to ensure the generalization of identified patterns (Feelders, Daniels, and Holsheimer, 2000). This theme encapsulates the process of uncovering intrinsic data properties (Peng et al., 2011) and establishing connections between items using various data representations (SAS Institute Inc, 2017).

**Visualizations** are key to capturing the intricacies of data. They enable the extraction of insights and recognition of patterns. By exhibiting data points and their interrelationships, visualizations provide insights that may not be obtained through tables or summary statistics (Brachman and Anand, 1996; Delen and Al-Hawamdeh, 2009; Microsoft, 2020; SAS Institute Inc, 2017). This is especially evident while exploring high-dimensional data, where coordinated visualizations reveal intricate data structures and distributions (Abbasi and H. Chen, 2008; Dietrich, 2016) by highlighting the most relevant information (K. Chen and Liu, 2006; Jackson, 2002). Interactive and iterative visual methods are essential for exploratory data analysis, as they facilitate direct engagement with the data, enabling deeper examination of relationships between variables and identification of hidden insights (Dutta and Bose, 2015; Fayyad, Piatetsky-Shapiro, and Smyth, 1996; Larson and Chang, 2016; Li, Thomas, and Osei-Bryson, 2016; Saltz, 2021). This process typically involves analyzing data subsets using statistical summaries and visual representations to evaluate data quality and distribution (Rollins, 2015). Visualization techniques, including projections of high-dimensional data in the two-dimensional space, can accelerate the discovery process by making it easier to identify clusters, outliers, and data gaps (Abbasi and H. Chen, 2008; K. Chen and Liu, 2006; Jackson, 2002; Zhuang, Wilkin, and Ceglowski, 2013).

## 4.5 Insights

Next, we present the tangible outcomes and insights obtained from analyzing the data. This includes specific deliverables such as reports and models, as well as the actionable insights that inform decision-making and subsequent modeling (see Figure 6).

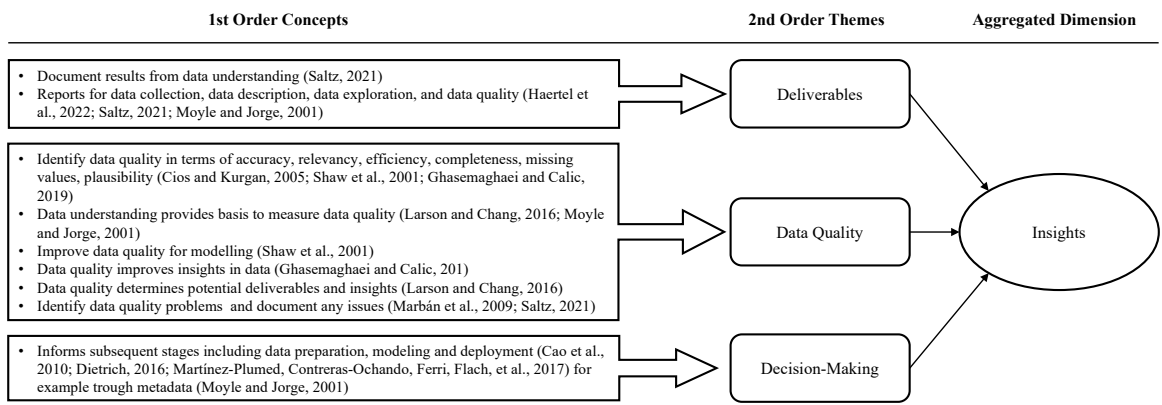

*Figure 6.*     *Data structure of insights*

**Deliverables** refer to the documentation produced as a result of the data understanding phase, which typically includes reports such as an initial data collection report, as noted in Moyle and Jorge (2001). This report outlines the specifics of the gathered data, providing a baseline for all subsequent analyses. It is followed by a data description report (Haertel et al., 2022; Moyle and Jorge, 2001; Saltz, 2021), which details the datasets' intrinsic characteristics. The data exploration report (Moyle and Jorge, 2001; Saltz, 2021) then captures insights and patterns uncovered during the exploration phase. The data quality report (Moyle and Jorge, 2001) complements this approach, assessing the data's reliability and appropriateness for analysis. Collectively, these reports offer a holistic view of the data, from initial collection to exploration and quality assessment, informing decision-making and strategic actions.

**Data Quality** describes the process of evaluating and assuring the integrity and usefulness of data. Essential tasks include the verification of data quality and the documentation of issues (Saltz, 2021), essential for developing and implementing data quality metrics (Larson and Chang, 2016), often informed by data profiling outcomes like demographics and descriptive statistics (Larson and Chang, 2016). Particularly with big data, challenges arise in maintaining accuracy and relevance due to the vastness of datasets (Ghasemaghaei and Calic, 2019; Martínez-Plumed et al., 2021). Enhancing data quality is crucial for modeling (Fatima et al., 2020) and involves evaluating the data's suitability for specific purposes, especially in relation to data mining and knowledge discovery goals (Cios and Kurgan, 2005). The accuracy and efficiency of data are fundamental in conducting thorough data analysis and generating valuable insights (Ghasemaghaei and Calic, 2019). The overall quality of data significantly influences the nature and depth of insights derived and the value of final deliverables (Larson and Chang, 2016).

**Decision-Making** underscores the significane of an in-depth data understanding for informed decision-making and effective application in later phases such as data preparation, modeling by turning insights from the data into tangible actions (Cao et al., 2010; Dietrich, 2016; Martínez-Plumed et al., 2017). The phase yields critical outputs like metadata and data quality information (Moyle and Jorge, 2001), which are integral to strategic decisions.

## 5 Conceptualizing Data Understanding

Through our literature review and inductive coding, we develop a framework based on the identified dimensions of data understanding, as shown in Figure 7. Before explaining the framework's dimensions and it's contribution to enhancing data understanding, we delineate it from similar concepts, such as business understanding (Wirth and Hipp, 2000) or data preparation (Quemy, 2020).

Business understanding involves comprehending the business context of the underlying problem, such as identifying appropriate performance metrics, determining required performance, or identifying stakeholders, i.e., determining what makes a project successful (Abbasi, Sarker, and Chiang, 2016). While the dimensions of our framework can inform business understanding, for example, by generating insights from linking data to the real-world, the focus is on understanding the data itself rather than the broader business context. Conversely, data preparation includes activities that transform the data based on the insights generated through data understanding. This can include, for example, engineering novel features, removing outliers, or other data-cleaning practices (Phillips-Wren et al., 2015). While several dimensions of our framework can inform these activities, they do not involve their actual execution. Instead, the dimensions provide the required foundational understanding to conduct them effectively. Further, specific methods, like data lineage, outlier identification, and interactive visualizations, exist that instantiate the identified themes and dimensions. For example, data lineage is part of the Foundations dimension, while outlier identification is an instantiation of Collection & Selection. Interactive visualizations, on the other hand, are a technique for informing the Exploration & Discovery of data. In contrast to these specific methods, our framework provides a broader perspective, integrating them into a holistic data understanding framework.

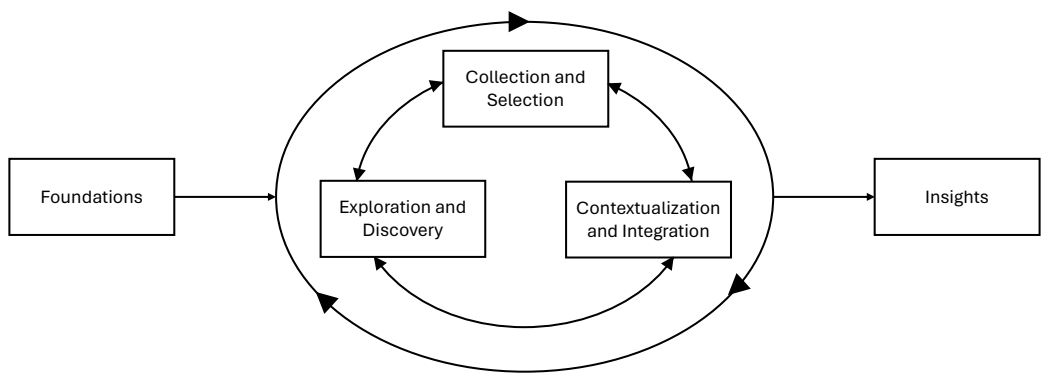

*Figure 7.    Data understanding consists of multiple interrelated dimensions*

The previously presented aggregated dimensions represent a logical order in which the essential phase of data understanding can be conducted. Starting with the **Foundations**, analysts search in data infrastructures like data warehouses or data lakes for relevant data. While doing so, they check the provenance of the discovered data to ensure that the transformations applied are valid and do not hinder effective analysis concerning the underlying use cases. To get an overview of the available data sources, they familiarize themselves and characterize the data through metadata and simple statistics.

The foundations then inform an iterative cycle that includes three central aggregated dimensions: **Collection & Selection**, **Contextualization & Integration**, and **Exploration & Discovery**. These interrelated elements form the core of the data understanding phase, each affecting and being affected by the others in a continuous loop of refinement and discovery. In Collection & Selection, the aim is to identify and gather relevant data sources informed by the initial foundational understanding. This helps to ensure that the data is comprehensive, relevant, and aligned with the analytical objectives. Contextualization & Integration then embed these selected data elements within specific domain contexts. This is essential for creating meaningful and actionable data, utilizing domain expertise to interpret and effectively integrate the data. Subsequently, Exploration & Discovery involve thoroughly examining the data and utilizing advanced analytical techniques to reveal hidden patterns, relationships, and insights. Yet these three parts iteratively and cyclically interact. For instance, findings obtained during the exploration and discovery may lead to the conclusion that additional or complementary data needs to be collected to address the underlying use case or that previously unknown relationships can be explained through domain experts. Similarly, as new data is acquired or the contextual landscape shifts, it is necessary to revisit the exploration and contextualize the data appropriately to uncover additional insights. This dynamic interplay ensures that the data understanding process is not linear but an iterative cycle of refinement and discovery, with each phase continuously informing and enhancing the others.

Ultimately, analysts want to generate **Insights** based on the acquired in-depth understanding of the data. Analysts must evaluate the data to determine its potential and decide whether to proceed with the project or abort it based on the quality of the data. They document their results in various reports, including data collection, description, exploration, and quality reports. If they decide to continue their project, they can use their understanding of the data to inform the following activities, like data preparation and modeling and, ultimately, the decision-making, thus leveraging the real-world value of the collected data.

# 6 Discussion and Conclusion

In this section, we discuss our findings on data understanding in analytics frameworks. We explore how our framework complements existing frameworks for analytics projects and it's position within the emerging paradigm of DCAI. Further, we explain the broader implications of our findings for both research and practice, reflect on the limitations of our study, and suggest avenues for future research.

**The presented framework complements existing frameworks to ensure a holistic and iterative approach to data understanding.** Current analytics frameworks often lack guidance on how to establish a comprehensive understanding of data to inform data preparation and modeling. Instead, they highlight different aspects, leaving a fragmented picture of what constitutes data understanding and, thus, untapped potential for tailoring data to the underlying analytic problem. To fill this gap, our proposed framework provides a holistic overview of the fundamental aspects that method experts (e.g., analysts) and domain experts (e.g., engineers) need to address. As such, it complements existing frameworks, like CRISP-DM, by presenting supplemental dimensions along which data needs to be understood. In contrast to many current frameworks, we emphasize the integral role that domain experts can play in understanding and contextualizing the data. This ensures that the data is technically sound while also reflecting the underlying domain correctly. However, this requires close collaboration between method and domain experts to contextualize the data appropriately (Holstein et al., 2023). The framework promotes collaboration to ensure a shared understanding and alignment of data insights with business objectives. This enhances the effectiveness and impact of data analytics initiatives. The framework contributes to advancing a more holistic, iterative, and collaborative approach to data understanding in the realm of data analytics.

**DCAI requires a more dynamic and comprehensive approach to data understanding.** In DCAI, the process of understanding data extends beyond the processes of traditional frameworks presented earlier. It involves a holistic evaluation of datasets to uncover and address underlying weaknesses. This involves pinpointing areas where data may be incomplete, incorrect, or biased and ensuring that datasets are not only technically sound but also contextually rich. Therefore, a key focus in DCAI is on systematically augmenting datasets. This goes beyond simply downsizing it to the most relevant data as often conducted in existing frameworks; instead, it involves actively identifying gaps and expanding the dataset to fill them, for example, by adding new labels or increasing the volume of highly relevant instances. Effectively expanding a dataset in this way necessitates an in-depth understanding of both the data and its domain, ensuring that the augmentations accurately represent the real-world scenarios they aim to model. Following the augmentation, an equally important aspect is the differentiation between outliers and informative edge cases. While most frameworks lack emphasis on this distinction, for DCAI, recognizing and leveraging these edge cases is critical. They provide valuable insights for building decision boundaries in AI models, thus increasing their performance. This shift to a more dynamic, comprehensive approach in DCAI underscores the need for frameworks to guide data understanding more effectively.

## 6.1    Implications for research and practice

Our findings provide a foundation for identifying the key dimensions that are essential for developing a solid understanding of data, thereby enhancing the theoretical knowledge of data understanding. By synthesizing existing approaches and inductively analyzing relevant research, we propose a framework for data understanding that complements current frameworks by extending them with guidance of relevant dimensions for data understanding. Future researchers can utilize these dimensions as the foundation for designing systems and methods that facilitate data understanding, offering practical implications for research and system development.

Our framework offers practical guidance by emphasizing the collaboration between method and domain experts and providing holistic guidance on essential dimensions for comprehensive data understanding. This collaboration ensures data analyses are both technically accurate and contextually relevant, bridging the gap between data science and domain expertise to generate business insights.

## 6.2    Limitations and future research

Our study, like any research, comes with certain limitations. One limitation is the framework's lack of specificity in recommending methods and techniques for practical application. While we outline key dimensions for understanding data, translating these conceptual guidelines into actionable steps remains a challenge. This gap points to the broader limitation of our study: The lack of empirical validation across different contexts and domains. Our framework is proposed based on a systematic literature review, without direct application or testing in real-world settings. Further, the framework emphasizes collaboration between method and domain experts without exploring the mechanisms or structures that facilitate this collaboration effectively. The assumption that such collaboration can be seamlessly integrated into existing processes may not hold in all organizational contexts.

Given the outlined limitations, future research should focus on two objectives. First, empirical studies should apply our framework across industries to evaluate its validity and adaptability, aiming to specify the methods, techniques, and tools used within its dimensions. This will help bridge the gap between conceptual understanding and actionable insights in operationalizing the framework. Second, research should examine structures that enable effective collaboration between method and domain experts, identifying best practices and solutions to enhance the framework's practicality and adoption, ensuring its approach to holistic data understanding can be effectively realized. In addressing these areas, future research can deepen our knowledge of data analytics by broadening our knowledge of data understanding. This will equip practitioners and scholars alike with the necessary guidance to develop DCAI solutions.

## 6.3    Conclusion

In this study, we perform a systematic literature analysis to identify the core elements of data understanding. We identify five fundamental dimensions and develop a framework to illustrate the aspects that need to be evaluated to establish a comprehensive understanding of data. Our analysis is contextualized in DCAI and its requirements for acquiring a profound insight into data to perform appropriate transformations and augmentations. Ultimately, the framework assists in developing a comprehensive understanding of data from the ever-expanding pool of available information.

## References

Aaltonen, A., Alaimo, C., Parmiggiani, E., Stelmaszak, M., Jarvenpaa, S., Kallinikos, J., and Monteiro, E. (2023). "What is Missing from Research on Data in Information Systems? Insights from the Inaugural Workshop on Data Research." *Communications of the Association for Information Systems* 53.

Abbasi, A. and Chen, H. (2008). "CyberGate: A Design Framework and System for Text Analysis of Computer-Mediated Communication." *MIS Quarterly* 32, 811–837.

Abbasi, A., Sarker, S., and Chiang, R. (2016). "Big Data Research in Information Systems: Toward an Inclusive Research Agenda." *Journal of the Association for Information Systems* 17, 1–32.

Brachman, R. J. and Anand, T. (1996). "The Process of Knowledge Discovery in Databases." In: *Advances in Knowledge Discovery and Data Mining*, pp. 37–57.

Cao, L. (2010). "Domain-driven data mining: Challenges and prospects." *IEEE Transactions on Knowledge and Data Engineering* 22 (6), 755–769.

— (2017). "Data Science: A Comprehensive Overview." *ACM Computing Surveys* 50 (3).

Cao, L., Zhao, Y., Zhang, H., Luo, D., Zhang, C., and Park, E. (2010). "Flexible Frameworks for Actionable Knowledge Discovery." *IEEE Transactions on Knowledge and Data Engineering* 22 (9), 1299–1312.

Chen, K. and Liu, L. (2006). "IVIBRATE: Interactive Visualization-Based Framework for Clustering Large Datasets." *ACM Transactions on Information Systems* 24 (2), 245–294.

Cios, K. J. and Kurgan, L. A. (2005). "Trends in data mining and knowledge discovery." In: *Advanced techniques in knowledge discovery and data mining*. Springer, pp. 1–26.

Cooper, H. (1988). "Organizing knowledge syntheses: A taxonomy of literature reviews." *Knowledge in Society* 1, 104–126.

Dag, A., Topuz, K., Oztekin, A., Bulur, S., and Megahed, F. M. (2016). "A probabilistic data-driven framework for scoring the preoperative recipient-donor heart transplant survival." *Decision Support Systems* 86, 1–12.

Delen, D. and Al-Hawamdeh, S. (2009). "A holistic framework for knowledge discovery and management." *Communications of the ACM* 52 (6), 141–145.

Dietrich, D. (2016). "Data analytics lifecycle processes." US9262493B1.

Dutta, D. and Bose, I. (2015). "Managing a Big Data project: The case of Ramco Cements Limited." *International Journal of Production Economics* 165, 293–306.

Fassnacht, M. K., Benz, C., Leimstoll, J., and Satzger, G. (2023). "Is Your Organization Ready to Share? A Framework of Beneficial Conditions for Data Sharing." *ICIS 2023 Proceedings*.

Fatima, F., Talib, R., Hanif, M. K., and Awais, M. (2020). "A Paradigm-Shifting From Domain-Driven Data Mining Frameworks to Process-Based Domain-Driven Data Mining-Actionable Knowledge Discovery Framework." *IEEE Access* 8, 210763–210774.

Fayyad, U., Piatetsky-Shapiro, G., and Smyth, P. (1996). "The KDD Process for Extracting Useful Knowledge from Volumes of Data." *Commun. ACM* 39 (11), 27–34.

Feelders, A., Daniels, H., and Holsheimer, M. (2000). "Methodological and practical aspects of data mining." *Information & Management* 37 (5), 271–281.

Gerhart, N., Torres, R., and Giddens, L. (2023). "Challenges in the Model Development Process: Discussions with Data Scientists." *Communications of the Association for Information Systems* 53.

Ghasemaghaei, M. and Calic, G. (2019). "Can big data improve firm decision quality? The role of data quality and data diagnosticity." *Decision Support Systems* 120, 38–49.

Gioia, D., Corley, K., and Hamilton, A. (2013). "Seeking Qualitative Rigor in Inductive Research." *Organizational Research Methods* 16, 15–31.

Guo, P. J. (2012). *Software tools to facilitate research programming*. Stanford University.

Haertel, C., Pohl, M., Nahhas, A., Staegemann, D., and Turowski, K. (2022). "Toward a lifecycle for data science: a literature review of data science process models." *PACIS 2022 Proceedings*.

Holstein, J., Schemmer, M., Jakubik, J., Vössing, M., and Satzger, G. (2023). "Sanitizing data for analysis: Designing systems for data understanding." *Electronic Markets* 33 (1), 52.

Hund, A., Wagner, H.-T., Beimborn, D., and Weitzel, T. (2021). "Digital innovation: Review and novel perspective." *The Journal of Strategic Information Systems* 30 (4), 101695.

Jackson, J. (2002). "Data mining; a conceptual overview." *Communications of the Association for Information Systems* 8 (1), 19.

Jakubik, J., Vössing, M., Kühl, N., Walk, J., and Satzger, G. (2024). "Data-Centric Artificial Intelligence." *Business & Information Systems Engineering*.

Jarrahi, M. H., Memariani, A., and Guha, S. (2023). "The Principles of Data-Centric AI." *Commun. ACM* 66 (8), 84–92.

Larson, D. and Chang, V. (2016). "A review and future direction of agile, business intelligence, analytics and data science." *International Journal of Information Management* 36 (5), 700–710.

Lebovitz, S., Levina, N., and Lifshitz-Assaf, H. (2021). "Is AI Ground Truth Really True? The Dangers of Training and Evaluating AI Tools Based on Experts' Know-What." *MIS Quarterly* 45, 1501–1526.

Li, Y., Thomas, M. A., and Osei-Bryson, K.-M. (2016). "A snail shell process model for knowledge discovery via data analytics." *Decision Support Systems* 91, 1–12.

Marbán, O., Segovia, J., Menasalvas, E., and Fernández-Baizán, C. (2009). "Toward data mining engineering: A software engineering approach." *Information Systems* 34, 87–107.

Mariscal, G., Marbán, O., and Fernández, C. (2010). "A survey of data mining and knowledge discovery process models and methodologies." *The Knowledge Engineering Review* 25, 137–166.

Martínez-Plumed, F., Contreras-Ochando, L., Ferri, C., Flach, P., Hernández-Orallo, J., Kull, M., Lachiche, N., and Ramírez-Quintana, M. J. (2017). "CASP-DM: context aware standard process for data mining." *arXiv*.

Martínez-Plumed, F., Contreras-Ochando, L., Ferri, C., Hernández-Orallo, J., Kull, M., Lachiche, N., Ramírez-Quintana, M. J., and Flach, P. (2021). "CRISP-DM Twenty Years Later: From Data Mining Processes to Data Science Trajectories." *IEEE Transactions on Knowledge and Data Engineering* 33 (8), 3048–3061.

Merkelbach, S., Von Enzberg, S., Kühn, A., and Dumitrescu, R. (2022). "Towards a Process Model to Enable Domain Experts to Become Citizen Data Scientists for Industrial Applications." In: *2022 IEEE 5th International Conference on Industrial Cyber-Physical Systems (ICPS)*, pp. 1–6.

Microsoft (2020). *Data acquisition and understanding of Team Data Science Process*. URL: `https://learn.microsoft.com/azure/architecture/data-science-process/lifecycle`.

Mikalef, P., Pappas, I. O., Krogstie, J., and Giannakos, M. (2018). "Big data analytics capabilities: a systematic literature review and research agenda." *Information systems and e-business management* 16, 547–578.

Moyle, S. and Jorge, A. (2001). "Ramsys-a methodology for supporting rapid remote collaborative data mining projects." In: *ECML/PKDD01 Workshop: Integrating Aspects of Data Mining, Decision Support and Meta-learning (IDDM-2001)*. Vol. 64.

Patel, H., Guttula, S., Gupta, N., Hans, S., Mittal, R. S., and N, L. (2023). "A data centric AI framework for automating exploratory data analysis and data quality tasks." *Journal of Data and Information Quality*.

Peng, Y., Zhang, Y., Tang, Y., and Li, S. (2011). "An incident information management framework based on data integration, data mining, and multi-criteria decision making." *Decision Support Systems* 51 (2).

Petricek, T., Den Burg, G. J. van, Nazábal, A., Ceritli, T., Jiménez-Ruiz, E., and Williams, C. K. (2022). "AI Assistants: A Framework for Semi-Automated Data Wrangling." *IEEE Transactions on Knowledge and Data Engineering*.

Phillips-Wren, G., Iyer, L. S., Kulkarni, U., and Ariyachandra, T. (2015). "Business analytics in the context of big data: A roadmap for research." *Communications of the Association for Information Systems*, 23.

Quemy, A. (2020). "Two-stage optimization for machine learning workflow." *Information Systems* 92, 101483.

Raftopoulos, M. and Hamari, J. (2023). "Human-ai collaboration in organisations: a literature review on enabling valure creation." In: *ECIS 2023 Research Papers*.

Rollins, J. B. (2015). *Foundational Methodology for Data Science*. Tech. rep. IBM.

Saltz, J. S. (2021). "CRISP-DM for Data Science: Strengths, Weaknesses and Potential Next Steps." In: *2021 IEEE International Conference on Big Data (Big Data)*, pp. 2337–2344.

Samtani, S., Zhu, H., Padmanabhan, B., Chai, Y., Chen, H., and Nunamaker, J. F. (2023). "Deep Learning for Information Systems Research." *Journal of Management Information Systems* 40 (1), 271–301.

SAS Institute Inc (2017). URL: `https : / / documentation . sas . com / doc / en / emref / 14 . 3 / n061bzurmej4j3n1jnj8bbjjm1a2.htm`.

Savarimuthu, B. T. R., Yasir, M., Corbett, J., and Lakshmi, V. (2023). "Improving Information Systems Sustainability by Applying Machine Learning to Detect and Reduce Data Waste." *Communications of the Association for Information Systems* 53.

Shaw, M. J., Subramaniam, C., Tan, G. W., and Welge, M. E. (2001). "Knowledge management and data mining for marketing." *Decision Support Systems*.

Snyder, H. (2019). "Literature review as a research methodology: An overview and guidelines." *Journal of Business Research* 104, 333–339.

Spitzer, P., Kühl, N., Heinz, D., and Satzger, G. (2023). "ML-Based Teaching Systems: A Conceptual Framework." *Computer Supported Cooperative Work and Social Computing*.

Webster, J. and Watson, R. T. (2002). "Analyzing the Past to Prepare for the Future: Writing a Literature Review." *MIS Quarterly* 26 (2), xiii–xxiii.

Whang, S. E., Roh, Y., Song, H., and Lee, J.-G. (2023). "Data collection and quality challenges in deep learning: a data-centric AI perspective." *The VLDB Journal* 32 (4), 791–813.

Wirth, R. and Hipp, J. (2000). "CRISP-DM: Towards a standard process model for data mining." *Proceedings of the 4th International Conference on the Practical Applications of Knowledge Discovery and Data Mining*.

Wolfswinkel, J., Furtmueller, E., and Wilderom, C. (2013). "Using Grounded Theory as a Method for Rigorously Reviewing Literature." *European Journal of Information Systems* 22.

Yu, L., Wang, S., and Lai, K. K. (2006). "An integrated data preparation scheme for neural network data analysis." *IEEE Transactions on Knowledge and Data Engineering* 18 (2), 217–230.

Zha, D., Bhat, Z. P., Lai, K.-H., Yang, F., and Hu, X. (2023a). "Data-centric AI: Perspectives and Challenges." In: *Proceedings of the 2023 SIAM International Conference on Data Mining (SDM)*. Proceedings. Society for Industrial and Applied Mathematics, pp. 945–948.

Zha, D., Bhat, Z. P., Lai, K.-H., Yang, F., Jiang, Z., Zhong, S., and Hu, X. (2023b). *Data-centric Artificial Intelligence: A Survey*.

Zhuang, Z. Y., Wilkin, C. L., and Ceglowski, A. (2013). "A framework for an intelligent decision support system: A case in pathology test ordering." *Decision Support Systems* 55 (2), 476–487.